\documentstyle[11pt,epsfig,aaspp4]{article}
\tighten 
\begin{document}

\newcommand{\lya}{Lyman~$\alpha$}
\newcommand{\lyb}{Lyman~$\beta$}
\newcommand{\za}{$z_{\rm abs}$}
\newcommand{\ze}{$z_{\rm em}$}
\newcommand{\cmtwo}{cm$^{-2}$}
\newcommand{\nhi}{$N$(H$^0$)}
\newcommand{\nzn}{$N$(Zn$^+$)}
\newcommand{\ncr}{$N$(Cr$^+$)}
\newcommand{\degpoint}{\mbox{$^\circ\mskip-7.0mu.\,$}}
\newcommand{\halpha}{\mbox{H$\alpha$}}
\newcommand{\hbeta}{\mbox{H$\beta$}}
\newcommand{\hgamma}{\mbox{H$\gamma$}}
\newcommand{\kms}{\,km~s$^{-1}$}      
\newcommand{\minpoint}{\mbox{$'\mskip-4.7mu.\mskip0.8mu$}}
\newcommand{\mv}{\mbox{$m_{_V}$}}
\newcommand{\Mv}{\mbox{$M_{_V}$}}
\newcommand{\peryr}{\mbox{$\>\rm yr^{-1}$}}
\newcommand{\secpoint}{\mbox{$''\mskip-7.6mu.\,$}}
\newcommand{\sqdeg}{\mbox{${\rm deg}^2$}}
\newcommand{\squig}{\sim\!\!}
\newcommand{\subsun}{\mbox{$_{\twelvesy\odot}$}}
\newcommand{\et}{et al.~}

\def\ltsima{$\; \buildrel < \over \sim \;$}
\def\simlt{\lower.5ex\hbox{\ltsima}}
\def\gtsima{$\; \buildrel > \over \sim \;$}
\def\simgt{\lower.5ex\hbox{\gtsima}}
\def\arcs{$''~$}
\def\arcm{$'~$}
\vspace*{0.1cm}
\title{Si AND Mn ABUNDANCES IN DAMPED LYMAN $\alpha$ 
SYSTEMS WITH LOW DUST CONTENT\altaffilmark{1}}

\vspace{1cm}
\author{\sc Max Pettini and Sara L. Ellison}
\affil{Institute of Astronomy, Madingley Road, Cambridge, CB3 0HA, UK}
\author{\sc Charles C. Steidel\altaffilmark{2} and Alice E. Shapley}
\affil{Palomar Observatory, Caltech 105--24, Pasadena, CA 91125}
\author{\sc David V. Bowen}
\affil{Princeton University Observatory, Princeton, NJ 08544}

\altaffiltext{1}{The data presented herein were obtained with the 
NASA/ESA {\it Hubble Space Telescope} and with the Keck~I Telescope.
The W.M. Keck Observatory is operated as a scientific partnership 
among the California Institute of Technology, the University of 
California, and the National Aeronautics and Space Administration. The 
Observatory was made possible by the generous financial support of the 
W.M. Keck Foundation.}
\altaffiltext{2}{NSF Young Investigator}

\newpage
\begin{abstract}
We have measured the abundances of Zn, Si, Mn, 
Cr, Fe, and Ni in three damped \lya\ systems at 
redshifts $z_{\rm abs} \leq 1$ from high resolution echelle spectra 
of QSOs recorded with the Keck~I telescope. 
In all three cases the  
abundances of Cr, Fe, and Ni relative to Zn indicate low levels of dust 
depletions. 
We propose that when the 
proportion of refractory elements locked up in dust grains is less than 
about 50\%, it is plausible to assume 
an approximately uniform level of depletion  
for all grain constituents and, by applying a small dust correction,
recover the intrisic abundances of Si and Mn.
We use this approach on a small sample of damped systems
(6 -- 8 cases) for which it is appropriate,
with the aim of comparing 
the metallicity dependence of the ratios [Si/Fe] and [Mn/Fe] 
with analogous measurements in Milky Way stars.
The main conclusion is that the relative
abundances of both elements in distant galaxies
are broadly in line with expectations based on Galactic data.
Si displays a mild enhancement at low metallicities, as expected for an 
$\alpha$-capture element, but there are also examples 
of near-solar [Si/Fe] at [Fe/H]~$ < -1$. 
The underabundance of Mn 
at low metallicities is possibly even more pronounced than that
in metal-poor stars, and no absorption system has yet been found where 
[Mn/Fe] is solar. 
The heterogeneous 
chemical properties of damped \lya\ systems, evident even from
this limited set of measurements, provide further support for
the conclusion from imaging studies that
a varied population of galaxies gives rise to 
this class of QSO absorbers.

We also present a {\it Hubble Space Telescope} 
image of the field of one 
of the QSOs, Q0058+019, showing the presence of an  
edge-on late-type galaxy only 1.2 arcseconds from the absorption sight-line. 
If this is the galaxy producing the damped \lya\ system at
$z_{\rm abs} = 0.61251$, it is of relatively low luminosity 
($M_B = -19.1$) and at an impact parameter of $10~h_{50}^{-1}$~kpc.
\end{abstract}

\newpage
\section{INTRODUCTION}

This is the fourth paper in a series dealing with metal abundances in 
damped \lya\ systems (DLAs) at intermediate redshifts 
($z_{\rm abs} < 1.5$). 
The number of such systems known has been increasing slowly 
over the past few years with the growing database of
{\it Hubble Space Telescope} (HST) QSOs 
observed at wavelengths below 3000~\AA\ which are 
inaccessible from the ground. 
In Pettini et al. (1999, Paper III)
we combined measurements of the abundance of Zn in 
10 DLAs at $z_{\rm abs} < 1.5$ 
with earlier surveys at higher redshifts 
to determine the evolution of the metallicity
of H~I gas in the universe.
We found that, somewhat surprisingly, the metal content of DLAs 
apparently does not increase with cosmic time, 
and that the column density-weighted mean value
of the Zn abundance remains roughly constant at
${\rm [} \langle{\rm Zn/H}\rangle {\rm ]} \simeq -1.1 \pm 0.2$
between $z = 3$ and 0.4\,.\footnote{In the usual notation, 
[Zn/H] = log (Zn/H) $-$ log (Zn/H)$_{\odot}$.}
This result is apparently at odds with the 
common interpretation of DLAs as the 
high redshift progenitors of present day spiral galaxies;
disk stars in the Milky Way, for example, had already reached 
[Fe/H]~$\approx -0.5$ 
at $z \approx 1$ (Edvardsson et al. 1993), 
a point first made by Meyer \& York (1992).
The persisting low metallicity 
of DLAs may be explained by abundance
gradients (Prantzos \& Silk 1998) but, more generally, 
is consistent with the finding from {\it HST} imaging
that galaxies of different morphological types and with a range of 
surface brightnesses contribute to the absorption cross-section for H~I
(Le Brun et al. 1997). 
To some extent this may well be a consequence of the fact that 
DLAs selected with {\it HST} are likely to be preferentially
dust- and therefore metal-poor, 
simply because the background QSOs would 
otherwise be too 
faint to be accessible with a 2.5~m telescope. 

Whatever the connection between them and present-day galaxies, 
damped \lya\ systems remain our best route to accurate determinations
of element abundances at high redshifts.
Element ratios in Galactic stars and nearby H~II regions
have long been scrutinized with a view to 
deciphering the clues they hold
both to the origin of different stellar populations and to 
the stellar yields (see, for example, Wheeler, Sneden, \& Truran 1989
for a review of the main ideas underlying this 
field of work).
As pointed out by Pettini, Lipman, \& Hunstead (1995),
abundance measurements in DLAs are potentially
an important extension of this technique 
(and one yet to be fully exploited),
allowing access to elements which are not well observed in stars,
to lower metallicities than those of present-day H~II regions, and
to a wider range of environments and physical conditions than 
local studies. A possible complication in interpreting interstellar
abundances is accounting for the fractions of refractory elements which 
are missing from the gas-phase having been incorporated into dust 
grains; however, we are aided in this respect 
by the generally low dust depletions 
which seem to apply to many DLAs (Pettini et al. 1997a).

In this paper we analyse Keck~I HIRES observations
of several elements, ranging from Mg to Zn, in three 
DLAs at $z_{\rm abs} \simlt 1$. One of these, 
at \za\ = 0.61251 in Q0058+019 (= PHL~938),
has not been studied before while for the other two,
at \za\ = 1.00945 in Q0302$-$223 and
\za\ = 0.85967 in Q0454+039, 
we previously published
only intermediate resolution observations 
(Pettini \& Bowen 1997, Paper II; Steidel et al. 1995, Paper I 
respectively).
In addition, we present an {\it HST} WFPC2 image of the 
field of Q0058+019 where we resolve a galaxy which is a highly 
plausible candidate for the damped absorber, 
being very close to the QSO sight-line.
We use the pattern of element abundances in these and other 
DLAs where corrections for dust depletion are estimated to be small
to explore the metallicity dependence of the abundances of 
Si (an $\alpha$-element)
and Mn. 


\section{{\it HST} OBSERVATIONS}
\subsection{FOS Spectroscopy}

A trawl of the {\it HST} Faint Object Spectrograph
(FOS) data archive
revealed that Q0058+019 exhibits a damped \lya\ line at
$\lambda_{\rm obs} = 1959$~\AA, 
shown in Figure 1.
In producing this spectrum we resampled the 
pipeline calibrated data to a linear dispersion
of 0.51~\AA\ per pixel (one quarter diode steps) and 
applied a correction of +8\% of the continuum level 
to bring the core of the \lya\ line to net
zero flux. A fit to the absorption profile
yielded a neutral hydrogen column density
$N$(H$^0$) = ($1.2 \pm 0.5$)$ \times 10^{20}$~cm$^{-2}$
at an absorption redshift \za\ = 0.6118. The column density
error, which includes the effect of the correction applied to the zero 
level, is larger than is usually the case because the signal-to-noise ratio
of the short FOS exposure is only $\approx 7$.  
Even so, the value of $N$(H$^0$) is lower than the 
threshold $N$(H$^0$) $ = 2 \times 10^{20}$~cm$^{-2}$ 
for DLAs originally
adopted by Wolfe et al. (1986), reflecting
the shift of the column density distribution toward lower
values at $z < 1.5$, as first noted 
by Lanzetta, Wolfe, \& Turnshek (1995).
The difference between the redshift of the \lya\ line and 
\za\ = 0.61251 measured from the metal absorption lines 
in the HIRES spectrum (\S3) 
corresponds to approximately half a diode 
on the detector and is typical 
of the accuracy with which the zero point 
of the FOS wavelength scale can be determined 
(Rosa, Kerber, \& Keyes 1998).

The FOS spectra of Q0302$-$223 and
Q0454+039 have been described in Papers II and I respectively; 
from a re-examination of the  
damped \lya\ line profiles we 
deduce $N$(H$^0$) = ($2.3 \pm 0.5$)$ \times 10^{20}$~cm$^{-2}$
and ($4.9 \pm 0.7$)$ \times 10^{20}$~cm$^{-2}$
at \za\ = 1.00945  and \za\ = 0.85967 respectively,
in good agreement with the values published in the earlier papers.

\subsection{WFPC2 Imaging}
The field of Q0058+019 was imaged with the Wide Field Planetary Camera
(WFPC2) as part of a larger {\it HST} program to study the morphology 
and environments of galaxies producing Mg~II absorption systems.
A set of four exposures was taken 
through the F702W filter (with an effective wavelength of 6900~\AA)
in a two-point dither pattern; the total exposure time was
5000~s. The individual CCD frames were
reduced using the pipeline calibration procedure and then
coadded by ``drizzling'' onto a master output pixel grid 
using the DITHER and 
DITHERII IRAF packages (Fruchter \& Hook 1999). 

The next step involved subtracting the QSO image to reveal any galaxies
at small separations. The {\it HST} Mg~II absorber imaging program
was purposefully designed to
facilitate this subtraction process by constructing
an empirical point spread function (PSF) 
using images of many QSOs.
Since the PSF characteristics (FWHM, shape, bleeding) depend
sensitively on the level of saturation, QSOs observed in
the program were grouped by flux so that
an appropriate PSF could be determined
using only QSOs of similar flux to Q0058+019.
Subtraction of this median PSF (with the DAOPHOT IRAF package)
then yielded the final image reproduced in Figure 2.

A faint galaxy is clearly visible approximately 1.2 arcsec to the 
north-east of the QSO. 
Given its proximity to the QSO sight-line
this is the most likely candidate for the 
damped \lya\ absorber at \za\ = 0.61251.
Apparently the model PSF does not reproduce accurately one of the
diffraction spikes in the QSO image, leaving a residual flux deficit 
which cuts through the galaxy. When this is taken into account,
the object morphology is suggestive of a late-type galaxy seen at a high 
inclination angle, $i \approx 65^{\circ}$.
Table 1 lists relevant measurements, assuming
$z_{\rm gal} = 0.61251$ and adopting a $H_0 = 50$~km~s$^{-1}$~Mpc$^{-1}$, 
$q_0 = 0.05$ cosmology. 
We converted the measured F702W magnitude 
to an AB magnitude in the ${\cal R}$ photometric system of
Steidel \& Hamilton (1993) by reference 
to ground-based images of the field (the [6930/1500] ${\cal R}$ filter is 
a very close match to the WFPC2 F702W filter), and obtained
${\cal R} = 23.7$. For the above cosmology this in turn corresponds to an 
absolute magnitude in the rest-frame $B$-band (in the conventional Vega-based
magnitude system) $M_B = -19.1$\,. No K-correction was applied because 
at $z = 0.61251$ 6930~\AA\ is close to the effective wavelength of the 
$B$-band.
Adopting $M_B^{\ast} = -21.0$ (e.g. Folkes et al. 1999),
we conclude that the candidate damped \lya\ absorber is a
galaxy of relatively low luminosity, with $L \simeq 1/6 L^{\ast}$.
Other DLAs have been found to be associated with compact galaxies, 
dwarfs, low surface brightness galaxies, and even an S0
(Le Brun et al. 1997; Lanzetta et al. 1997; Rao \& 
Turnshek 1998). We now add a low luminosity spiral to the list
and thereby reinforce the conclusion of these earlier studies
that damped \lya\ systems are drawn from a diverse 
population of galaxies with a wide range of morphologies and  
luminosities.\\

\section{KECK OBSERVATIONS}

The spectra of the three QSOs were recorded 
at high spectral resolution with the HIRES echelle
spectrograph (Vogt et al. 1994)
on the Keck I telescope on Mauna Kea, Hawaii on
23 and 24 September 1998. 
Relevant details of the observations are
collected in Table 2. 
We used the UV-blazed cross-disperser grating
to record interstellar  
lines of interest in the three DLAs
longward of the atmospheric cut-off near 3200~\AA.
Thus, in Q0058+019 
we cover from Zn~II~$\lambda 2026$
to Mg~I~$\lambda 2852$ at \za\ = 0.61251,
while the spectra of Q0302$-$223 and Q0454+039
extend from Si~II~$\lambda 1808$ to Mn~II~$\lambda 2606$
at \za\ = 1.00945 and 0.85967 respectively (for
Q0302$-$223 this necessitated a second grating setting).
Given the good seeing (0.5 to 0.7 arcsec), we used the 
0.86~arcsec wide entrance slit which projects to 3 pixels on the
2048x2048 Tektronix CCD detector resulting in a resolution of 6.5
km~s$^{-1}$~FWHM.

The echelle spectra were extracted with Tom Barlow's customised
software package, following the steps described in Paper III.
The signal-to-noise ratios of the reduced spectra were measured directly 
from the rms fluctuations about the continuum level.
In general the value of S/N varies along each spectrum due to the 
presence of broad emission lines at the QSO redshifts and increasing 
atmospheric absorption below 3600~\AA; the values (per pixel) listed in 
column (10) of Table 2 refer to the region near rest frame wavelength  
$\lambda_0 = 2350$~\AA\ and should be representative of most of the 
absorption lines recorded. As can be seen from column (11) our 
Keck spectra are sensitive to rest frame equivalent 
widths of only a few m\AA. Figures 3, 4, and 5 show examples of 
absorption lines of varying strengths in each damped system. \\

\section{ION COLUMN DENSITIES AND ELEMENT ABUNDANCES}

When recorded at high spectral resolution the metal lines
in QSO absorption systems 
commonly break up into multiple components;
as can be seen from Figures 3, 4, and 5,
the three DLAs observed here are no exception with the strongest
absorption lines extending over 
100 to 200~km~s$^{-1}$. 
We analysed these complex absorption profiles with the VPFIT software 
written by Bob Carswell. The procedure has been described in detail 
before (e.g Paper III). As our data include seven transitions 
of Fe~II with widely different $f$-values, spanning a range of $\sim 165$,
the model fits produced with VPFIT are well constrained and allow us
to determine the redshift, velocity
dispersion parameter {\it b} ($b = \sqrt{2} \sigma$, where
$\sigma$ is the one-dimensional velocity dispersion of the
ions along the line of sight, assumed to be Gaussian),
and ion column density {\it N} of each component.
Details of the profile fits are collected in Table~3.
The important point is that the total ion column densities
do not depend on the fine detail of the profile decomposition because 
for each species considered we observe sufficiently weak transitions 
that the corresponding absorption lines fall on the linear part of the 
curve of growth. The exception is the Mg~II~$\lambda\lambda 2796,2803$ 
doublet which is strongly saturated (see Figure 3) and is therefore not 
included in the present analysis.

The total column densities of the first ions of 
Zn, Si, Mn, Cr, Fe, and Ni
in each DLA are listed in Table 4, together with \nhi. 
In deriving these values we used the 
compilation of $f$-values by
Morton (1991) with the revisions proposed by Savage \& Sembach (1996).
For Ni~II we took advantage of the recent radiative lifetime measurements
by Fedchack \& Lawler (1999) which have led to a reduction by a 
factor of 1.9 of the $f$-values of the $\lambda 1709.600$ and $\lambda 
1741.549$ transitions relative to the values proposed by Zsarg\'{o} \& 
Federman (1998). The ensuing upward revision of 
$N$(Ni$^+$) by a factor of $\simeq 2$ is significant for the 
interpretation of the pattern of relative element abundances, as 
discussed below (\S5). 

Since in DLAs the elements considered here
are predominantly singly ionized,
their abundances can be deduced
directly by dividing the values of {\it N} in columns (4) to (9)
by the values of \nhi\ in column (3) of Table 4. Comparison with
the solar system abundance scale of Anders \& Grevesse (1989) finally
gives the relative abundances listed in Table 5.
(We have included in Table 5 the abundance measurements from
Paper~III, with the appropriate revisions for $N$(Ni$^+$), 
as they will be considered in the discussion below.)   
We now briefly describe the results for the three 
DLAs which are the subject of the present paper.

1. Q0058+019; $z_{\rm abs} = 0.61251$: Among the species
observed in our HIRES spectra Zn gives the most direct measure of 
metallicity, free from the complication of dust depletion.
In this low redshift DLA the Zn~II~$\lambda2026,2062$ doublet lines
fall at 3267 and 3326~\AA\ respectively, where observations are difficult 
due to atmospheric absorption. Although noisy, we clearly detect both 
lines; given the relatively low column density of hydrogen in this 
system, the presence of Zn~II absorption in itself implies high 
abundances and indeed we deduce [Zn/H] $\approx 0$.
The conclusion that this absorber has near-solar metallicity does not 
rest on the poorly observed Zn~II lines alone; 
as can be seen from Table 5, the abundances of Cr and Fe, two other 
iron-peak elements, are also within a factor of $\sim 2$ of solar and 
could be higher if some fraction of these elements has been incorporated 
into dust grains, as discussed below (\S5). 

Evidently, DLAs with near-solar abundances are not rare at 
$z \simlt 1$; out of the six measurements available to date,
three have metallicities $Z_{\rm DLA} \simgt 1/3 Z_{\odot}$
(see Figure 7 of Paper III). 
However, a wide range of values of [Zn/H], spanning $\sim 1.5$~dex,
persists at all redshifts. 
It is intriguing that  
systems with high metallicity
are invariably at the low end of the distribution 
of neutral hydrogen column density, 
so that the census of metals seen in absorption is dominated by 
gas with high \nhi\ and low metal 
content.\footnote{Thus, the new measurement for this DLA 
has a minimal effect on the column-density weighted average
${\rm [} \langle{\rm Zn/H}\rangle {\rm ]} = -1.03 \pm 0.23$
in the redshift interval $z = 0.40 - 1.5$ derived in Paper III.}
It is highly likely that selection effects play a role here; 
DLAs with large columns of molecules (and therefore probably high 
metallicity) are known to exist
(e.g. the \za\ = 0.68466 21~cm absorber towards 
B0218+357---Carilli, Rupen, \& Yanny 1993; Wiklind \& Combes 1995), 
but are too faint to be studied spectroscopically 
at optical and ultraviolet wavelengths. 
It is a lingering concern, 
however, that such selection effects 
are still largely unquantified.

Galaxies in the local universe exhibit a rough correlation between
metallicity and $B$-band luminosity which apparently persists at least to 
$z = 0.4$ (Kobulnicky \& Zaritsky 1999). Referring to these authors' 
Figure 4 it can be seen that, 
with $M_B = -19.1$ and [Zn/H]~$=+0.1 \pm 0.2$,
the \za\ = 0.61251 absorbing galaxy in Q0058+019 is somewhat 
metal-rich for its luminosity but is not inconsistent with the local 
relationship given the observed scatter. What is perhaps more remarkable
is to find a near-solar abundance at relatively 
large distances from the centre 
of the galaxy. If the DLA arises in the disk,
the high inclination of the 
galaxy, $i \approx 65^{\circ}$,
places it at a galactocentric distance of 
$10~h_{50}^{-1}$/cos~$i \simeq 24~h_{50}^{-1}$~kpc. 
If the absorption takes place in the 
halo, it would imply the existence of a cloud with 
$N$(H~I)~$= 10^{20}$~cm$^{-2}$ and solar metallicity 
$10~h_{50}^{-1}$~kpc above the 
mid-plane of the galaxy.
In either case it would seem that this galaxy does not have
a marked abundance gradient, either along or perpendicular to the disk.

2. Q0302$-$223; $z_{\rm abs} = 1.00945$:  As can be seen from Figure 4,
two main groups of components, 
separated by 36~km~s$^{-1}$, produce most of the absorption 
seen in this DLA; additional weaker components, 
at $v = +35$ and $+121$~km~s$^{-1}$ 
relative to \za\ = 1.00945 are visible in the stronger Fe~II lines.
Although the Zn~II and Cr~II lines are weak, the corresponding column 
densities and abundances in Tables 3 and 4 are in excellent agreement
with the values reported in Paper II which were measured from 
data of much lower resolution (0.88~\AA\ compared to 
0.08~\AA~ FWHM). {\it HST} WFPC2 imaging (Le Brun et al. 1997) has revealed two 
compact galaxies close to the line of sight. 
At $z = 1.009$ they would have absolute luminosities 
$L_B \approx 0.2 L_B^{\ast}$ and $\approx L_B^{\ast}$
and impact parameters $12~h_{50}^{-1}$ and $27~h_{50}^{-1}$~kpc 
respectively. It remains to be established with spectroscopic 
observations which of the two galaxies is associated with the damped 
absorber.

3. Q0454+039; $z_{\rm abs} = 0.85967$: Two groups of components, 
separated by $70$~km~s$^{-1}$, are responsible for most of the 
absorption in this DLA (Figure 5). Again, the column densities  
we deduce for Zn$^+$, Cr$^+$, and Fe$^+$ are in excellent agreement
with the values measured  
by Steidel et al. (1995) from 2.3~\AA\ resolution Lick spectra, 
once allowance is made for the different $f$-values used.
These authors also reported the presence of a compact galaxy close to the 
line of sight to the QSO, subsequently confirmed with WFPC2 images by Le 
Brun et al. (1997). If this is the absorber, it is at a projected 
separation of $8 h_{50}^{-1}$~kpc and it has an absolute luminosity
$L_B \approx 0.25 L_B^{\ast}$. The low element abundances we find, 
approximately 1/10 of solar, indicate that this galaxy apparently does 
not conform to the metallicity-luminosity relation 
discussed by Kobulnicky \& Zaritsky (1999); in this respect it
is more in line with present-day H~II galaxies.\\

\section{DUST DEPLETIONS}

The pattern of relative abundances measured in the 
interstellar gas of distant galaxies responds to two
effects, the selective depletion of refractory elements
onto dust grains and inherent differences from the solar system
scale, reflecting the past history of star formation which may well have 
been different from that of the Milky Way disk. 
Our goal here is to separate these two effects and, by accounting for the 
first, gain an insight into the second.
In this endeavour we are guided by the extensive body of data
on element abundances in stellar populations and the interstellar medium 
(ISM) of our Galaxy. 

Of the elements considered here, Zn, Cr, Fe, and Ni all track each other 
closely in Galactic stars with metallicities $-2.0 \leq {\rm [Fe/H]} \leq 0.0$
(e.g. Ryan, Norris, \& Beers 1996). 
Dust depletion, on the other hand, is more significant for Cr, Fe, and Ni 
than for the generally undepleted Zn (Savage \& Sembach 1996).
It follows from these considerations that we can take the ratios
[Zn/Cr], [Zn/Fe], and [Zn/Ni] as indicative of the fractions of these 
refractory elements which are missing from the gas-phase, e.g.
\begin{equation}
{\rm [Zn/Cr]} = {\rm log}~ 
\left ( \frac{f_{\rm gas} + f_{\rm dust}}{f_{\rm gas}} \right )
\end{equation}
where $f_{\rm gas}$ and $f_{\rm dust}$ are the fractions of Cr in gaseous 
and solid form respectively.\footnote{Recently, Howk \& Sembach (1999)
have drawn attention to the fact that ionization effects may 
boost the $N$(Zn$^+$)/$N$(Cr$^+$) ratio thereby mimicking dust depletion.
However, this is unlikely to be the case for the DLAs considered here.
Such effects would also increase the $N$(Ni$^+$)/$N$(Cr$^+$) ratio 
by similar factors, contrary to our measurements.
Based on the calculations by Howk \& Sembach (1999),
the observed [Ni/Cr]~$\simlt 0.0$ implies very low ionization parameters, 
as expected for clouds with $N$(H~I)$\simgt 10^{20}$~cm$^{-2}$.}

The boxes with the heavy outline in Figure 6 show the abundances measured
in the six intermediate redshift DLAs listed in Table 5.
Two conclusions can be drawn from inspection of this Figure. 
First, the depletions of Cr, Fe and, when available, 
Ni are roughly comparable, as 
is the case in the local ISM (Savage \& Sembach 1996).\footnote{The 
revision in the oscillator strengths of the Ni~II lines mentioned in 
\S4 has brought this element into better agreement with Fe and Cr, 
resolving an apparent puzzle which had been noted by Lu et al. (1996)
and in Paper III.} Second, the depletion levels  
are relatively modest, 
ranging from near-zero in Q0454+039 
to a factor of $\sim 4$ 
in Q1351+318 ($\sim 25$\% of Cr, Fe, and Ni in the gas).
Such values are typical of DLAs in general (Pettini et al. 1997a), 
whereas in the disk of the Milky Way the same elements 
are depleted by more than one order of magnitude 
(see Figure 6 of Savage \& Sembach 1996).

It is unclear what lies at the root of this difference. It is 
interesting that the ISM of the Small Magellanic Cloud also exhibits only 
mild depletions (Welty et al. 1997), but metallicity alone is unlikely to 
be the explanation, because there is no trend in our data for a 
dependence of [Zn/Cr] on [Zn/H] (for example 
Cr and Fe are depleted by only a factor of $\sim 2$ in Q0058+019, where 
[Zn/H] is approximately solar). In any case, it appears that in most DLAs
the balance between the incorporation of refractory elements into, 
and their release from, dust grains is shifted relative to the 
physical conditions prevailing  
in cool disk clouds on the Milky Way, so that on average 
there are roughly equal proportions of these elements 
in gas and dust. Note that this is unlikely to be 
the result of dust-related 
selection effects analogous to those mentioned earlier
(\S4). The total column densities of metals in the DLAs studied here are 
too low to produce significant dust reddening, even if 100\% of the 
elements which make up the grains were in solid form.

Finally on this topic, we point out that in two cases, Q0302$-$223 
and Q0454+039, we can determine depletions separately for the two 
well resolved groups of components which make up the absorption lines
(see Figures 4 and 5).
The results are summarised in Table 6.
Predictably, in Q0454+039 both components appear to be dust-free (since 
their sum is!). In Q0302$-$223, however, we see that the gas with 
the higher optical depth, at the adopted systemic redshift \za\ = 1.00945,
has [Zn/Cr,~Fe,~Ni]~$\simeq +0.6$, while in the component at
$v = -36$~km~s$^{-1}$ the same ratio is only $\simeq +0.1$. 
Reduced depletions in interstellar clouds with high  
velocities are commonplace in the local ISM, where they have been known for 
nearly 50 years (Routly \& Spitzer 1952) and are understood to arise from 
grain destruction in interstellar shocks.\\

\section{ELEMENT RATIOS}

Two of the elements covered by our observations, Si and Mn,
exhibit metallicity dependent ratios (relative to Fe) 
in Galactic stars, presumably because their nucleosynthesis  
follows different channels from that of the Fe-peak group.
Furthermore, in the local ISM both elements show a degree of dust depletion.
When overall depletion levels are high, 
Mn and Si are normally less underabundant
than Fe, Cr, and Ni. However, such differences become less 
pronounced as depletions are reduced 
(Figure 6 of Savage \& Sembach 1996)---all 
elements tend to the same depletion as the overall depletion level 
approaches zero. If we restrict ourselves
to cases where [Zn/Cr]~$\simlt 0.3$ 
(and therefore $f_{\rm dust} \simlt f_{\rm gas}$ in eq. (1) so that
dust correction factors are $\simlt 2$), 
we may be justified in assuming that to a first 
approximation all refractory elements are depleted by the same 
factor.\footnote{In future it should be possible to test this assumption 
by measuring the abundance of S, an undepleted $\alpha$-element.
If the assumption is correct, we expect [S/Si] = [Zn/Cr,~Fe,~Ni].}
The boxes with the light outline in Figure 6 show element abundances
corrected for the dust fractions implied by the observed [Zn/Cr] ratios; 
in each case adopting the mean [Zn/Cr,~Fe,~Ni] ratio 
would produce very similar results. 
We take these values to represent 
the total abundances (gas + dust) of the element concerned
and, having made this correction, we can now proceed to compare the 
abundances of Si and Mn in DLAs of 
different metallicities with analogous measurements in  
Galactic stars.

The approach taken here is similar to, but more conservative
than, the analysis by Vladilo (1998) who also used the ratio of
Zn to Fe-peak elements to correct for dust depletion. The main
difference is in the fact that Vladilo applied the correction to
all DLAs for which relevant measurements were available,
irrespectively of the degree of depletion, with the assumption
that dust in DLAs has the same composition as in the Milky Way ISM.
In our opinion we are on safer ground by limiting ourselves to
cases where $f_{\rm dust} \simlt f_{\rm gas}$ because our
conclusions do not then depend sensitively on the unknown
detailed make-up of interstellar dust at high redshift.

\subsection{Silicon}
The data for Si are displayed in Figure 7a, where the 
dots are the stellar 
measurements by Edvardsson et al. (1993)
and Nissen \& Schuster (1997).
The general trend is for a mild increase 
in the relative 
abundance of Si at low metallicity;
[Si/Fe]~$\approx +0.2$ to $+0.3$ at 
[Fe/H]~$\simlt -1$\,.
This is an example of 
the well known overabundance of the $\alpha$-elements
which is generally attributed to the delayed production 
of additional Fe by Type Ia supernovae. 
In this picture, the overall metallicity (as measured by [Fe/H])
at which the ratio [$\alpha$/Fe] begins to decline towards the solar value
depends on the previous history of star formation.
A galaxy which turns most of its gas into stars within $\sim 1$~Gyr
(the generally assumed timescale for the explosion of Type Ia supernovae)
would maintain an enhanced [$\alpha$/Fe] ratio while [Fe/H] grows
to high values.
Such a scenario may apply to the thick disk and the bulge
of the Milky Way where recent observations seem to indicate 
a uniform enhancement of
the $\alpha$-elements at all metallicities
(Fuhrmann 1998; Rich 1999, private communication).
At the other extreme, in a galaxy where star formation proceeds slowly, 
or in bursts separated by 
quiescent periods lasting more than 1~Gyr,
there would be time for [$\alpha$/Fe] to decline to the solar value
(or even lower) while [Fe/H] remains low (Gilmore \& Wyse 1991;
Pagel \& Tautvaisviene 1998).

Returning to Figure 7a, we now consider the evidence provided by DLAs.
Triangles show our measurements from Table 5, corrected for dust as 
explained above and taking Zn as a proxy for Fe.\footnote{This approach 
is preferable to using our Fe abundances directly, because the latter are 
based on very weak transitions with oscillator strengths
which may be less secure than those of the Zn~II doublet. The systematic 
underabundance by  $0.2$~dex of Fe relative to Cr in Figure 6 may well
reflect the relative uncertainty in the $f$-values of the Cr~II and Fe~II 
lines.} We also searched the literature for other DLAs where
the abundances of Zn, Cr, and Si have been measured and the 
ratio [Zn/Cr] is within a factor of two of solar
(therefore implying correspondingly small dust corrections).
We found four such cases, 
all from the recent compilation by Prochaska \& 
Wolfe (1999); 
they are shown in Figure 7a as large filled dots, again with the 
assumption that 
Cr and Si are depleted by similar amounts.
The most straightforward conclusion from Figure 7a is that 
the [Si/Fe] ratio in DLAs is not dissimilar from the values observed in 
Galactic stars. At least half of the DLA measurements fall well in line 
with the bulk of stellar data. 
There are also hints of differences, with
two or three cases where
[Si/Fe] appears to be approximately solar at [Fe/H]~$\simlt -1$.
While it is premature to make too much of these differences, 
it is probably fair to say that,
unlike the situation for Galactic stars,
one would not discern any trend in the 
[Si/Fe] ratio with metallicity from the DLA results alone.
Thus, the data available at present are certainly 
consistent with the view that 
damped \lya\ absorbers are drawn from a varied population of galaxies 
which may have processed their interstellar gas at different rates 
prior to the time when we observe them.  
On the other hand, blanket statements to the effect that 
the chemical histories of DLAs 
are different from that of the Milky Way (e.g. Vladilo 1998)
do not seem to be fully justified on the basis of the data 
for Si in Figure 7a.

\subsection{Manganese}

A major new study of the abundance of Mn
has recently been completed by 
Nissen et al. (2000) who measured [Mn/Fe] in 119 Galactic 
F and G stars from the thin disk, the thick disk, and the halo,
following the same method as Edvardsson et al. (1993)
and making use of Hipparcos parallaxes where available.
The analysis takes into account hyperfine structure splitting of the 
Mn~I lines, which is one of the complications involved in bringing 
together different data sets from previous studies.
The measurements by Nissen et al. are reproduced as small dots
in Figure 7b. There is an obvious drop
in [Mn/Fe] with decreasing [Fe/H]; in the most metal-poor disk stars,
at [Fe/H]~$\simeq -1$, [Mn/Fe]~$\simeq -0.4$\,.
The physical processes responsible for the metallicity dependence 
of [Mn/Fe] have not yet been confidently identified.
Observers have remarked on the fact that the [Mn/Fe] trend seems to
mirror the overabundance of the $\alpha$-elements but in the opposite 
sense (e.g. McWilliam 1997), 
leading to the conjecture that Type Ia supernovae
may be an important source of Mn 
(Samland 1998; Nakamura et al. 1999).
On the other hand, nucleosynthesis calculations
can reproduce the shape of the trend in Figure 7b with a metallicity 
dependence yield of Mn in massive stars which in the calculations by 
Timmes, Woosley, \& Weaver (1995) overwhelms the Type Ia contribution.

The filled triangles in Figure 7b
show the values of [Mn/Fe] determined for five of the DLAs 
considered in this paper, again taking Zn as the proxy for Fe for the 
reason explained above. From a literature search 
we found only one other data point 
which could be included in our analysis,
in the \za\ = 1.3726 DLA towards Q0935+417 where [Zn/H]~$= -0.80$,
[Cr/H]~$= -0.90$, and [Mn/H]~$= -1.48$ (Meyer, Lanzetta, \& Wolfe 1995;
Pettini et al. 1997b).
The comparison between stars and DLAs
is complicated by the fact that 
there is still some uncertainty regarding the 
correct solar system value of the abundance of Mn.
Anders \& Grevesse (1989) 
quote log~(Mn/H)~$= -6.61$ in the solar photosphere, but 
log~(Mn/H)~$= -6.47$ from meteorites 
(the latter value is the one used in the present 
analysis). This discrepancy persists in the more recent
reappraisal of `Standard Abundances' by Grevesse, Noels, \& Sauval (1996).
The uncertainty in (Mn/H)$_{\odot}$ does not 
affect the stellar data of Nissen et al. (2000) 
which are all derived from differential 
measurements, but it does mean that there is a 0.14~dex 
ambiguity in referring the DLA values to the stellar scale.
Thus the triangles and filled large dot 
in Figure 7b may need to be raised by 0.14~dex should 
the meteoritic abundance determination turn out to be in error.

Even with this caveat, it does appear that Mn 
is underabundant in the galaxies producing damped \lya\ systems
by factors similar to those measured in Galactic metal-poor stars.
As in the case of Si, there are no obvious trends from the QSO absorption 
line data alone;
[Mn/Fe]~$\simeq -0.4 \pm 0.1$ 
is an adequate description of the whole DLA 
sample available at present.
It is intriguing that the underabundance of Mn seems to persists
to metallicities as high as solar, although admittedly such a statement 
is at present based on only one measurement.
If further cases are found in future, 
the hypothesis that the underabundance of Mn is due to a
metallicity-dependent yield in massive stars would clearly run into 
difficulties. On the other hand, 
finding that Mn is low
([Mn/Fe] $= -0.51$) in one DLA (in Q1354+258)
which shows no enhancement of Si at low metallicity 
([Fe/H] $= -1.61$---see Figure 7), 
argues against the SN Type Ia interpretation,
as also pointed out by Nissen et al. (2000).
Possibly a third process, yet to be identified, is responsible for the 
metallicity dependence of the abundance of Mn.

\section{CONCLUSIONS}

We have measured element abundances 
in three galaxies which give rise to damped \lya\ systems 
at intermediate redshifts ($z \simeq 0.61 -1.01$).
The new data confirm the well established result
that significantly smaller fractions of refractory elements are 
incorporated into dust grains in DLAs 
compared with interstellar clouds of similar column 
density in the disk of the Milky Way. 
Although the physical reasons underlying this effect
are not fully understood, 
empirically it appears
that the equilibrium between gas and dust in 
damped absorbers is shifted
so that on average comparable proportions
of the grain constituents are in gaseous and solid forms.
We propose that in cases where dust depletions are less 
than a factor of about two,
it is possible to account for the unobserved fractions of
Si, Mn, Cr, Fe, and Ni 
by assuming that they are all depleted by 
approximately the same factor.
This assumption then allows us to examine the 
dependence on metallicity 
of the intrinsic abundances of Si and Mn.

We find that the abundances of both elements
are broadly in line with values measured 
in metal-poor stars of the Milky Way.
In about half of the cases considered 
Si is mildly enhanced relative to Fe-peak elements
at the typically lower-than-solar metallicities of the DLAs,
but there are also counterexamples where [Si/Fe] is more nearly solar 
even though [Fe/H] is less than 1/10 solar. 
The underabundance of Mn at low metallicities is possibly even more 
pronounced than in Galactic stars, and no DLA has yet been found with a 
solar [Mn/Fe]. However, for neither element is there a 
clear abundance trend with 
metallicity; in our view this is an indication 
that galaxies picked by 
damped \lya\ absorption 
have experienced a variety of star formation histories
prior to the time when we observe them.
In this respect chemical abundances give a picture
consistent with the 
results from imaging studies (including new observations reported here) 
which have shown that galaxies 
associated with DLAs exhibit a wide range of morphologies, luminosities, and 
surface brightnesses. 

It is important to emphasize the preliminary nature of these conclusions
which are based on the comparison of very few measurements in DLAs
with a much larger body of stellar data. One of the lessons from stellar  
work is that there is considerable scatter, observational and intrinsic,
in the relative abundances of different elements so that most 
trends only become apparent when a large set of observations
has been assembled. 
As a field of study, abundance determinations in high redshift galaxies 
are some twenty years behind their counterparts in
Galactic stars but they may well hold the key to clarifying some of the 
still unresolved issues on the origin of elements.\\
  
\acknowledgements

It is a pleasure to acknowledge the competent assistance with the
observations by the staff of the Keck Observatory; special thanks are due 
to Tom Barlow for generously providing his echelle extraction software. 
We are very grateful to Poul Nissen and YuQin Chen for allowing us to use
their stellar Mn abundances in advance of publication, 
and to Jim Lawler and Steve Federman for the early communication
of their measurements of the $f$-values of Ni~II transitions.
This work has benefited from several conversations with
colleagues, particularly Ken'ichi Nomoto, 
Bernard Pagel, Jason X. Prochaska, and Sean Ryan.
C. C. S. acknowledges support from the National Science
Foundation through grant AST~94-57446 and from the David and Lucile 
Packard Foundation.


%
%

\begin{deluxetable}{lcccccc}
\tablewidth{17cm}
\tablecaption{PROPERTIES OF THE CANDIDATE DLA ABSORBER IN Q0058+019\tablenotemark{a}}
\tablehead{
\colhead{$\Delta\alpha$($''$)} &
\colhead{$\Delta\delta$($''$)} & \colhead{$\theta$($''$)}
& \colhead{$d$\tablenotemark{b,c} ~~~(kpc)} 
& \colhead{$R_s$ (mag)} & \colhead{$M_B$ (mag)\tablenotemark{c}} & \colhead {Comments}
}
\startdata
0.8   &0.85 & 1.2 & 10.3  & 23.7 & $-19.1$ & Edge-on, late-type spiral \\
\enddata
\tablenotetext{a}{Assuming the galaxy to be at the DLA redshift 
$z_{\rm abs} = 0.61251$}
\tablenotetext{b}{Projected separation from QSO sight-line}
\tablenotetext{c} {$H_0 = 50$~km~s$^{-1}$~Mpc$^{-1}$; $q_0 = 0.05$}
\end{deluxetable}
\newpage

%
%

\begin{figure}
\figurenum{0}
\psfig{figure=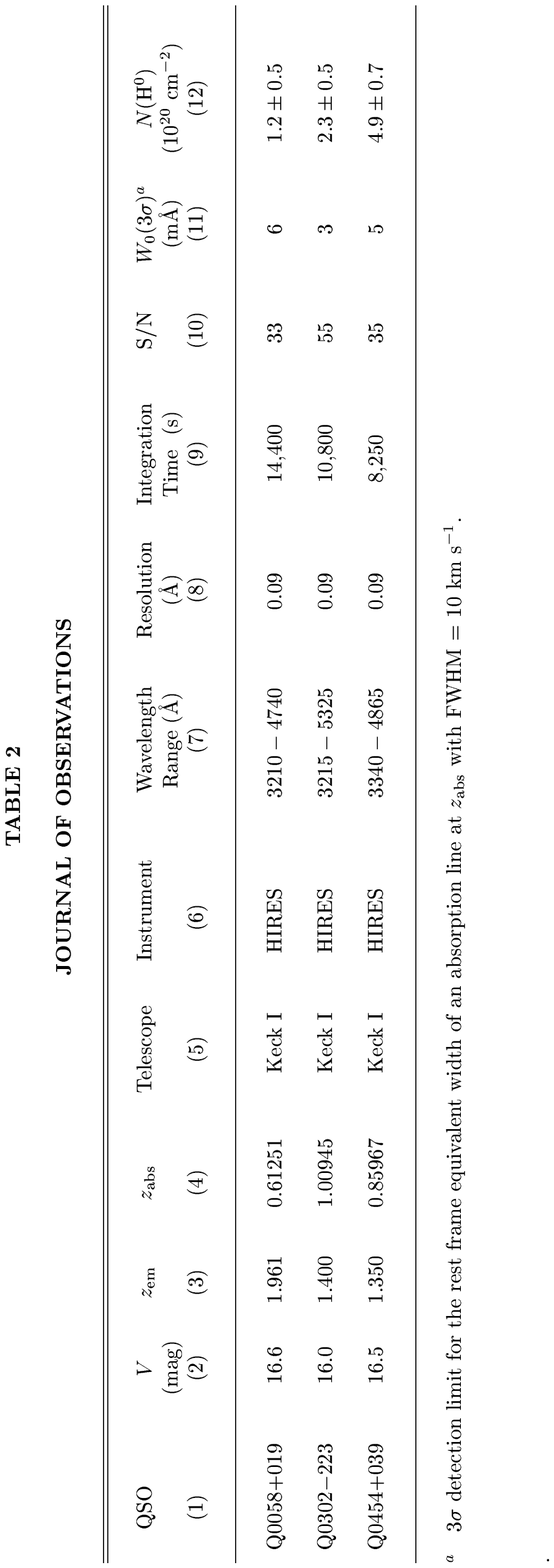,width=170mm,angle=180}
\end{figure}

\newpage

%
%

\addtocounter{table}{+1}
\begin{deluxetable}{lcccc}
\tablewidth{15cm}
\tablecaption{COMPONENT STRUCTURE IN THE Fe~II LINES}
\tablehead{
\colhead{QSO} &
\colhead{Component No.} & \colhead{$z_{\rm abs}$}
& \colhead{$b$ (km~s$^{-1}$)} & \colhead{log $N$\/(Fe$^+$) (cm$^{-2}$)}
}
\startdata
Q0058+019    &1 & 0.612240 & 6.0  & 12.83 \\
             &2 & 0.612508 & 6.1  & 14.77 \\
             &3 & 0.612616 & 24.9 & 15.01 \\
             &4 & 0.612866 & 8.0  & 13.69 \\
             &5 & 0.612940 & 2.4  & 13.25 \\
             &6 & 0.613041 & 16.8 & 13.28 \\
             &7 & 0.613139 & 7.4  & 13.63 \\
\\
Q0302$-$223  &1 & 1.009025 & 4.6  & 11.97 \\
             &2 & 1.009177 & 10.2 & 13.92 \\
             &3 & 1.009244 & 4.0  & 13.86 \\
             &4 & 1.009397 & 8.9  & 14.01 \\
             &5 & 1.009452 & 9.7  & 14.29 \\
             &6 & 1.009566 & 2.6  & 12.89 \\
             &7 & 1.009685 & 2.9  & 12.90 \\
             &8 & 1.010266 & 7.6  & 12.13 \\
\\
Q0454+039    &1 & 0.859225 & 5.4  & 14.38 \\
             &2 & 0.859248 & 16.8 & 14.00 \\
             &3 & 0.859401 & 3.9  & 13.17 \\
             &4 & 0.859486 & 6.1  & 13.78 \\
             &5 & 0.859670 & 14.8 & 15.00 \\
             &6 & 0.859917 & 9.4  & 13.83 \\
             &7 & 0.860045 & 6.2  & 12.22 \\
\enddata
\end{deluxetable}

\newpage

%
%

\begin{figure}
\figurenum{0}
\vspace*{-2cm}
\hspace*{-4.0cm}
\psfig{figure=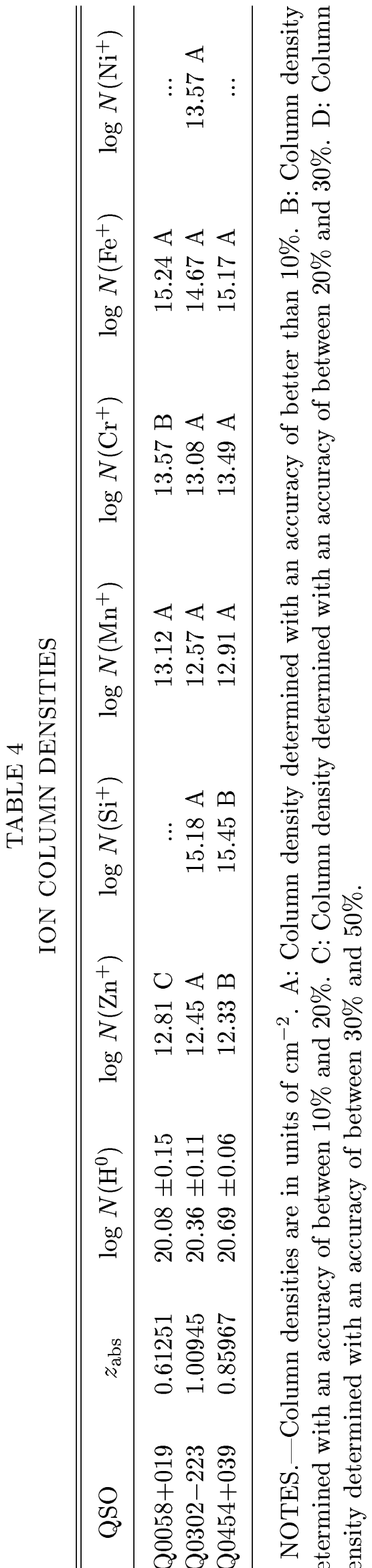,width=170mm,angle=180}
\end{figure}

\newpage

%
%

\begin{figure}
\figurenum{0}
\vspace*{-2cm}
\hspace*{-3cm}
\psfig{figure=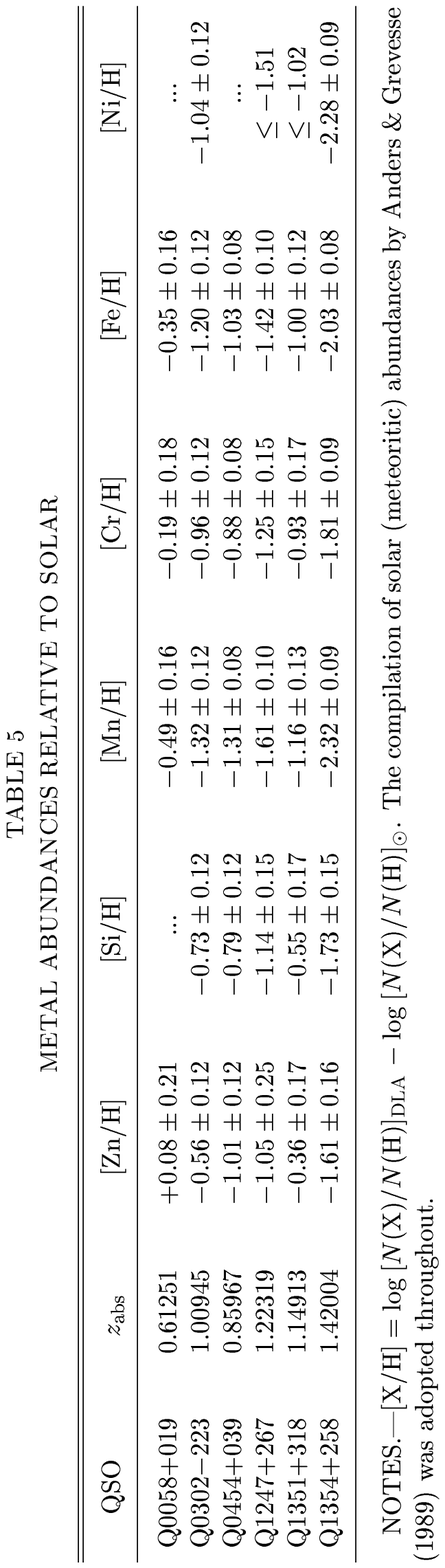,width=170mm,angle=180}
\end{figure}

\newpage

%
%

\begin{figure}
\figurenum{0}
\vspace*{-2cm}
\hspace*{-3cm}
\psfig{figure=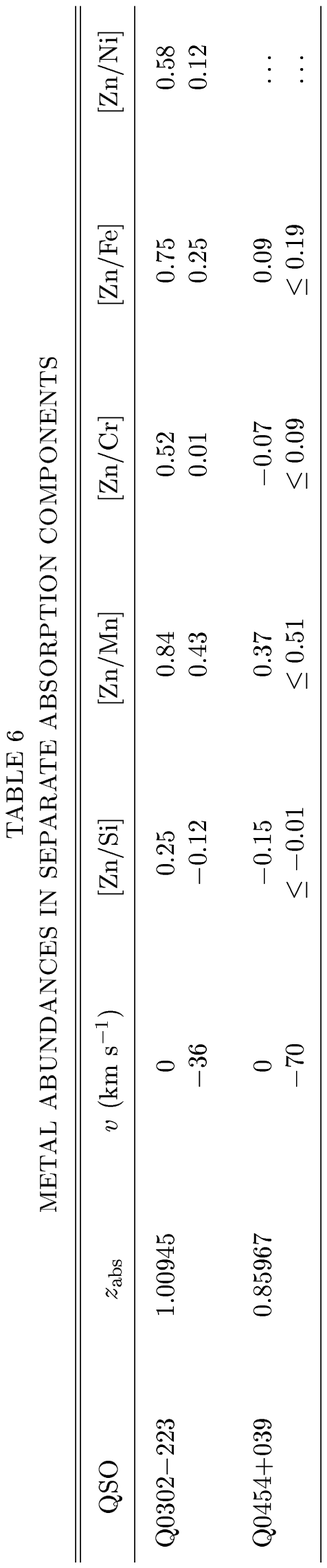,width=170mm,angle=180}
\end{figure}

\vfill

%
%

%
%

\begin{figure}
\figurenum{1}
\vspace*{-2cm}
\hspace*{-4.0cm}
\psfig{figure=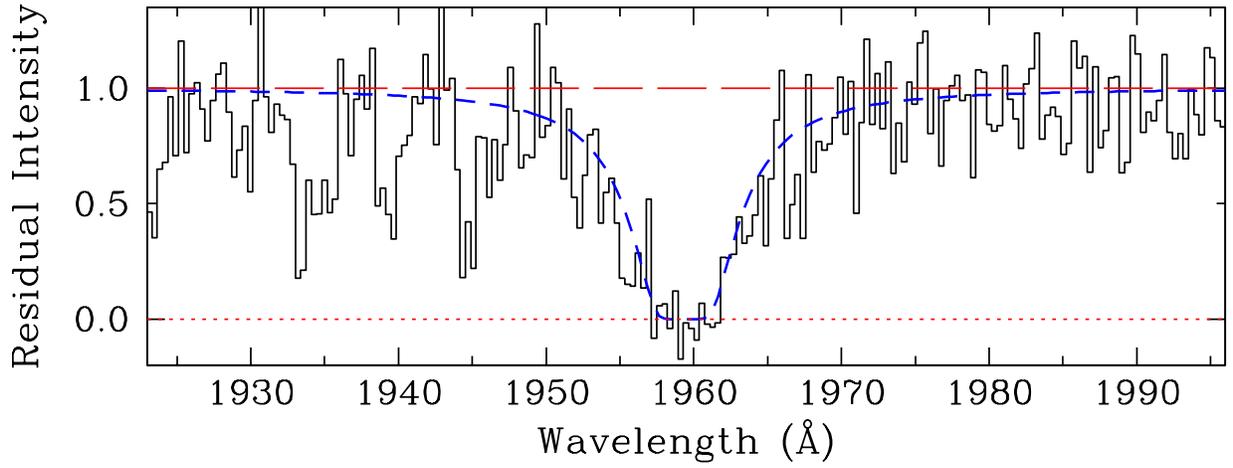,width=195mm,angle=270}
\vspace{-2.5cm}
\figcaption{Portion of the G270H FOS spectrum of
Q0058+019 (PHL~938) encompassing the
damped \lya\ line in the $z_{\rm abs} = 0.61251$ system.
The resolution is $\sim 2$~\AA\ FWHM and S/N $\approx  7$; the exposure 
time was 1590~s.
The short-dash line shows the theoretical damping profile
for a column density
$N$(H$^0$)$~= 1.2 \times 10^{20}$~cm$^{-2}$\,. }
\end{figure}

%
%

\begin{figure}
\figurenum{2}
\hspace*{-0.5cm}
\psfig{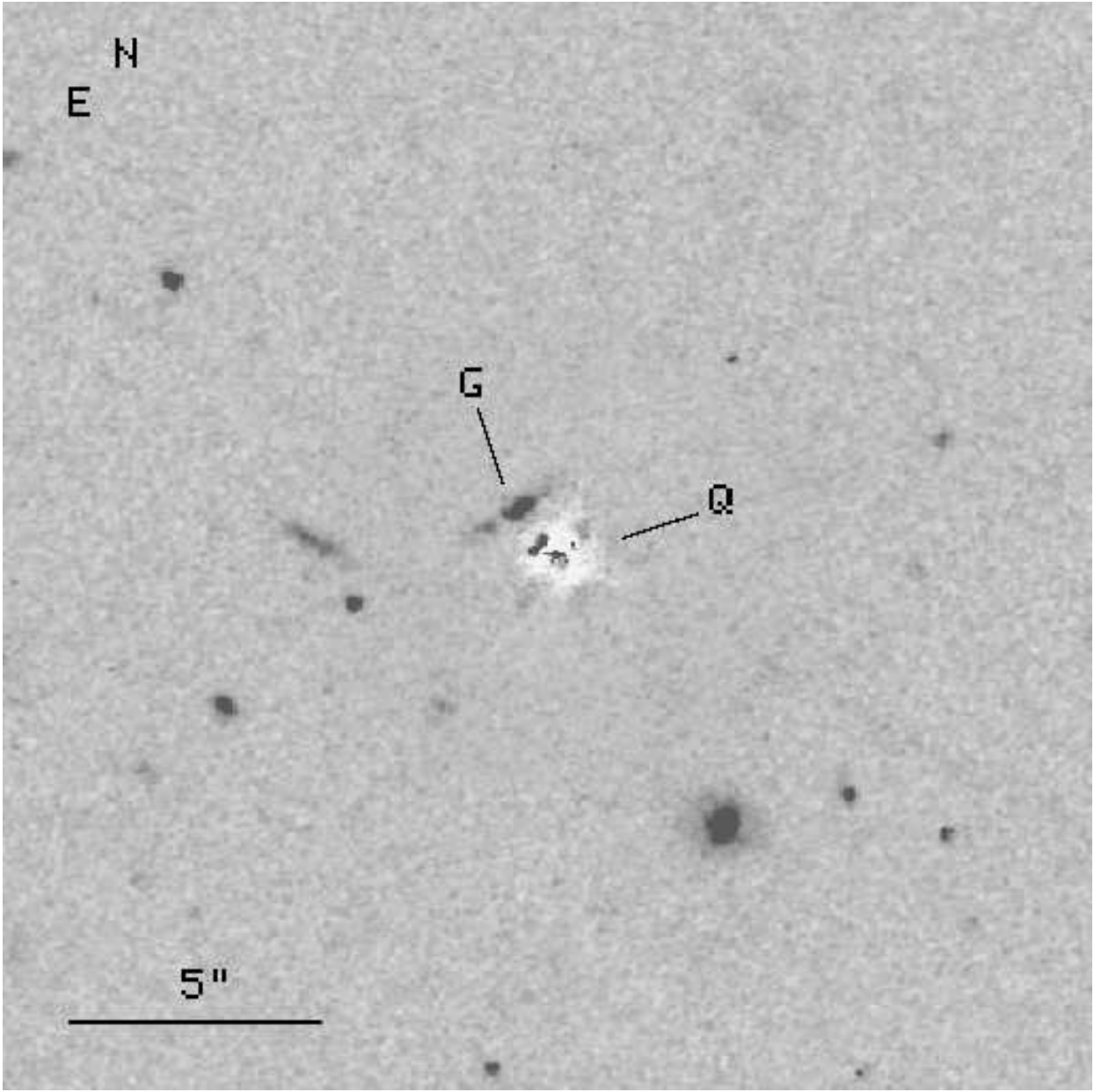}
\vspace{1cm}
\figcaption{WFPC2 F702W exposure of the field of Q0058+019 (PHL~938). North 
is at the top and East is to the left. A model point spread function has 
been subtracted from the QSO image as described in the text, 
revealing the presence of a galaxy approximately 1.2 arcseconds to the NE 
of the QSO position. 
Given its proximity, this is likely to be the damped 
\lya\ absorber at $z = 0.61251$. 
Residual excess absorption of a diffraction spike cuts across the
galaxy image. When this processing artifact is taken
into account, the candidate absorber appears to be a low
luminosity ($L \simeq 1/6L^*$) late-type galaxy seen at high
inclination, $i \approx 65^{\circ}$, at a projected
separation of $10~h_{50}^{-1}$~kpc from the QSO sight-line.
Other relevant measurement are collected in 
Table 2.}
\end{figure}

%
%

\begin{figure}
\figurenum{3}
\vspace*{-1.5cm}
\hspace*{-1cm}
\psfig{figure=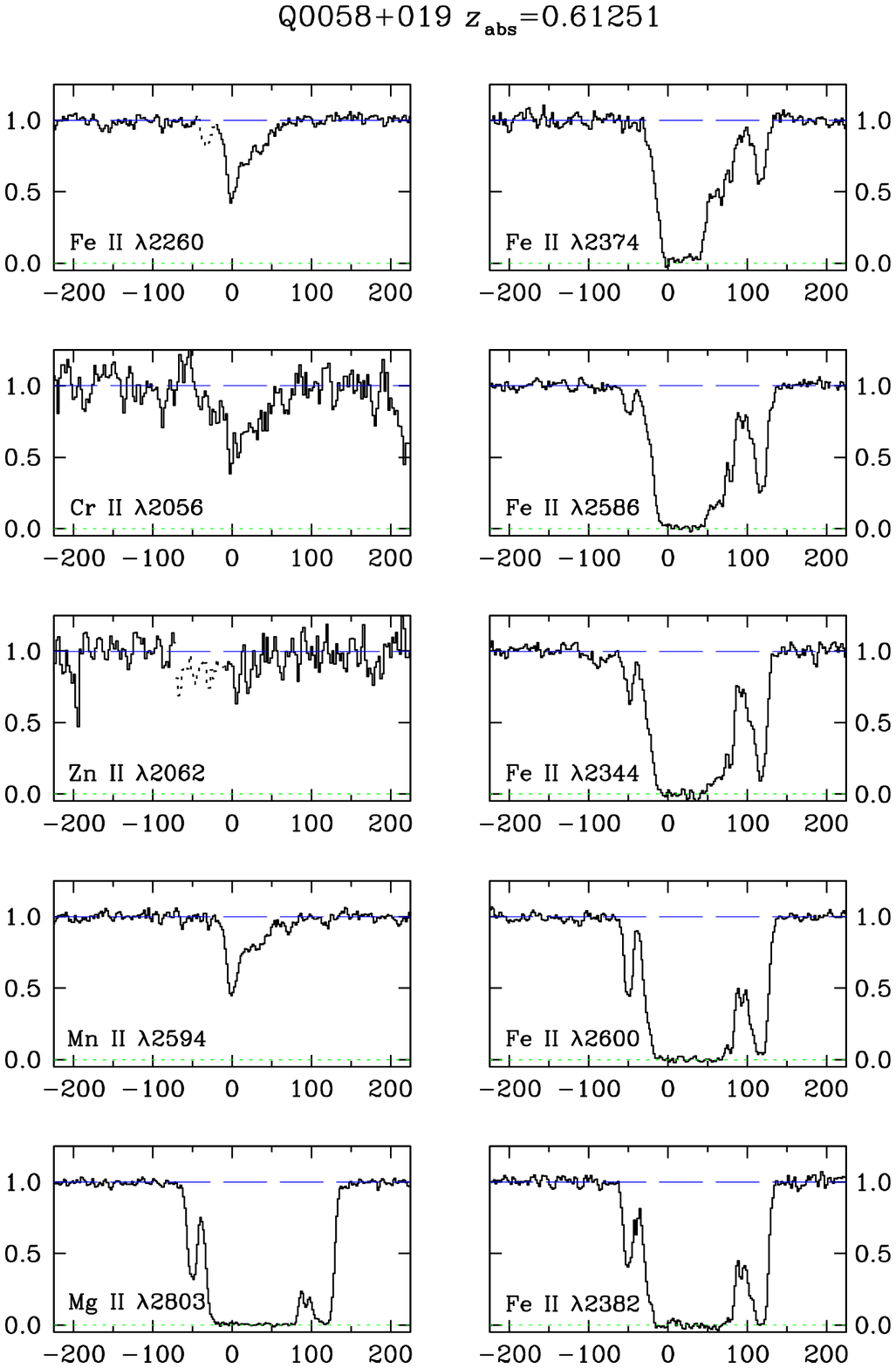,width=185mm}
\vspace{-4.5cm}
\figcaption{Profiles of selected absorption lines in the
\za\ = 0.61251 DLA in Q0058+019. The $y$-axis is residual intensity; the
$x$-axis is velocity in km~s$^{-1}$ relative to \za, defined from the
centroid of the component with the highest optical depth.}
\end{figure}

%
%

\begin{figure}
\figurenum{4}
\vspace*{-1.5cm}
\hspace*{-1cm}
\psfig{figure=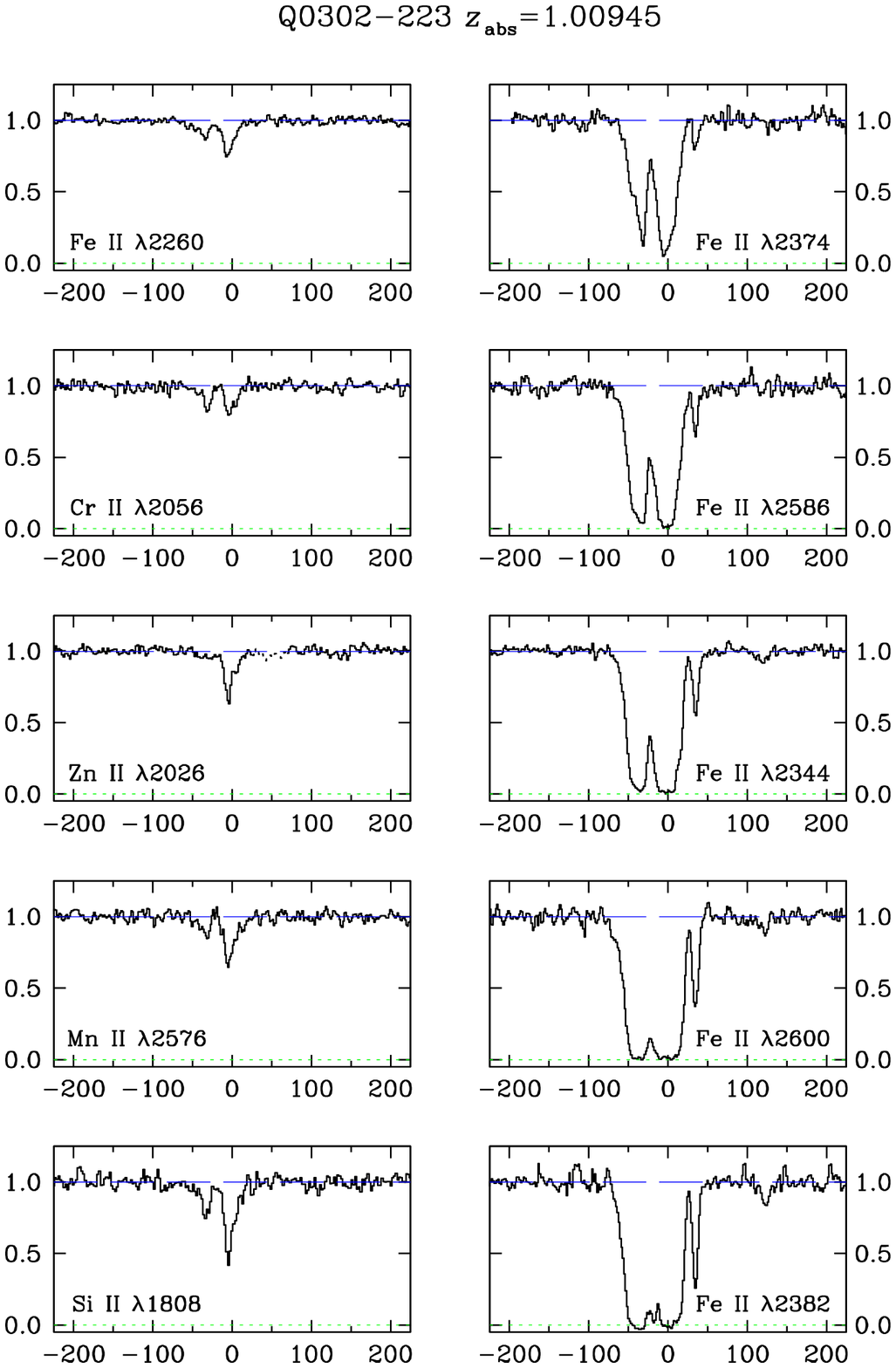,width=185mm}
\vspace{-4.5cm}
\figcaption{Profiles of selected absorption lines in the
\za\ = 1.00945 DLA in Q0302$-$223. The $y$-axis is residual intensity; the
$x$-axis is velocity in km~s$^{-1}$ relative to \za, defined from the
centroid of the component with the highest optical depth.}
\end{figure}

%
%

\begin{figure}
\figurenum{5}
\vspace*{-1.5cm}
\hspace*{-1cm}
\psfig{figure=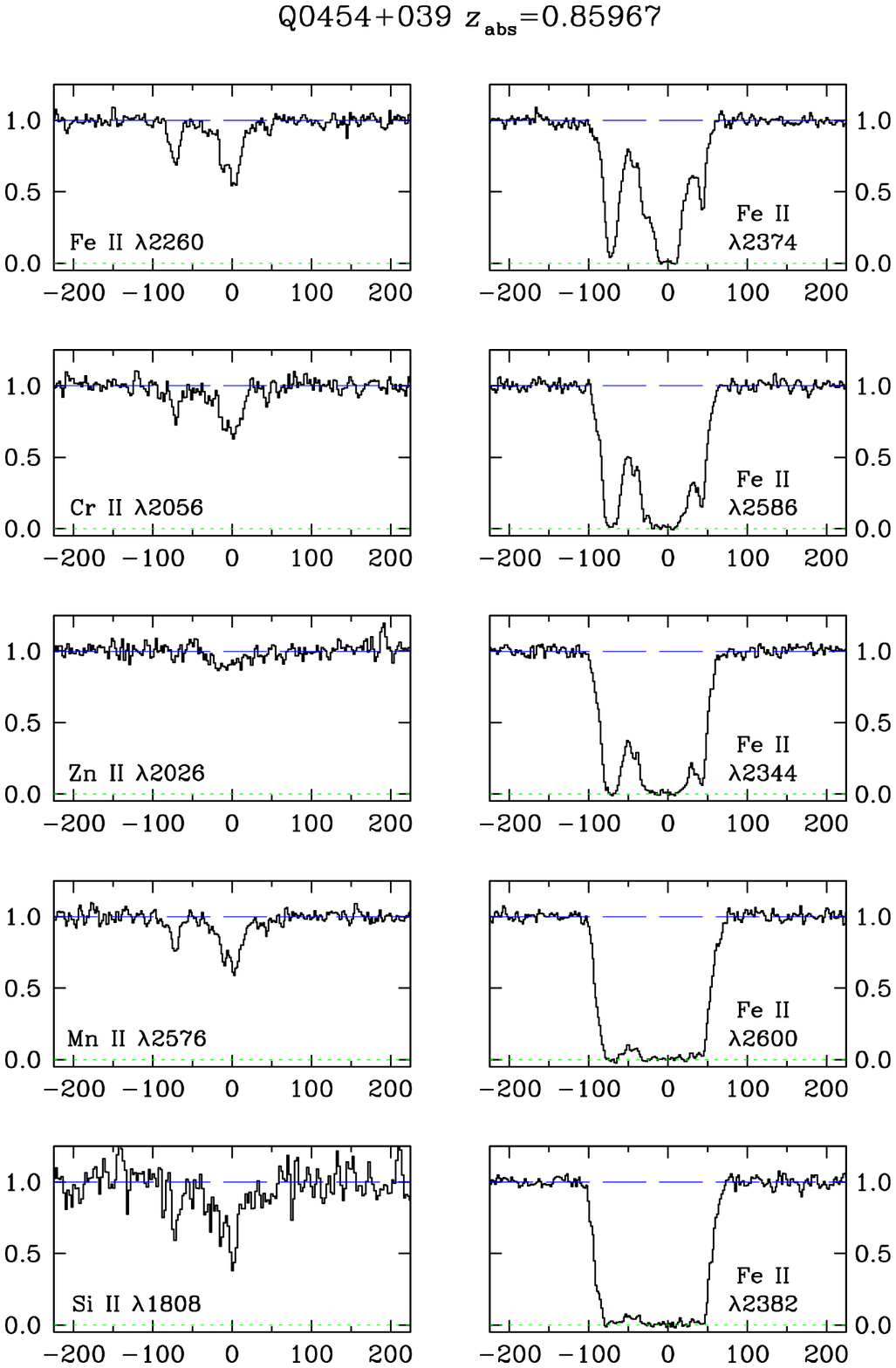,width=185mm}
\vspace{-4.5cm}
\figcaption{Profiles of selected absorption lines in the
\za\ = 0.85967 DLA in Q0454+039. The $y$-axis is residual intensity; the
$x$-axis is velocity in km~s$^{-1}$ relative to \za, defined from the
centroid of the component with the highest optical depth.}
\end{figure}

%
%

\begin{figure}
\figurenum{6}
\hspace*{-0.5cm}
\psfig{figure=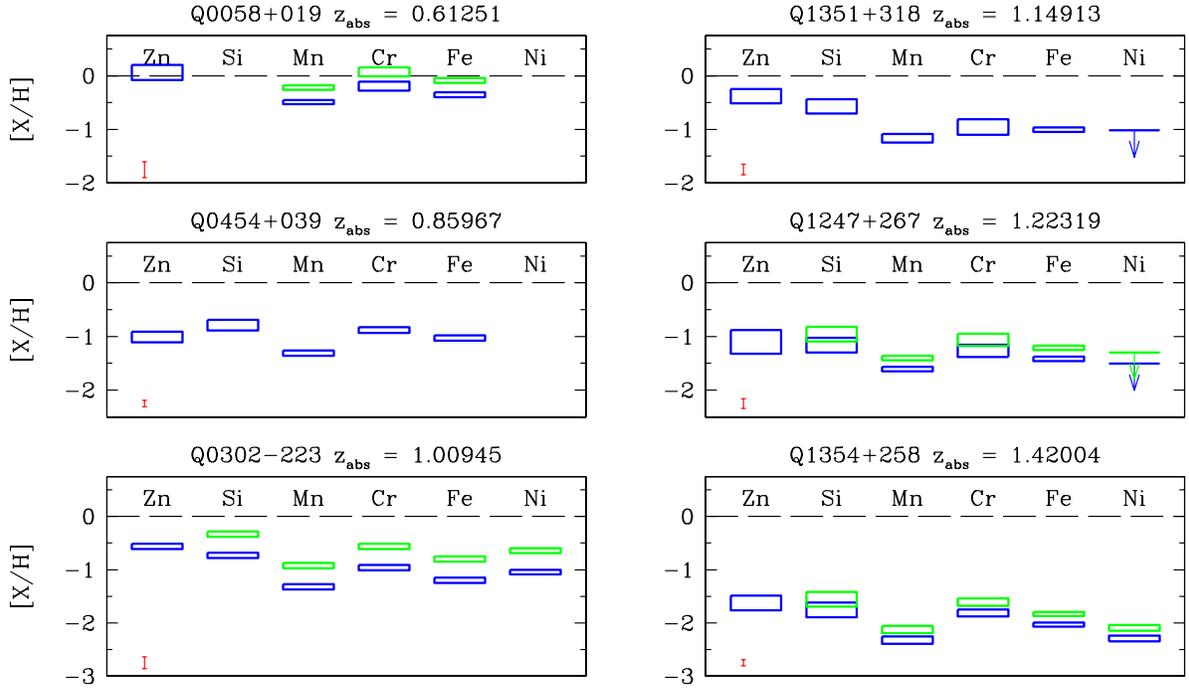,width=130mm,angle=270}
\vspace{-1cm}
\figcaption{Element abundances in intermediate redshift
DLAs from this paper and from Paper III are plotted on a logarithmic
scale relative to solar values. The height of each box represents
the uncertainty in the column density of that element; the
vertical bars near the bottom left hand corners of the panels
indicate the errors in the column densities of neutral hydrogen.
Boxes with heavy outline (blue in the colored version of this figure)
show the measured values; boxes with a light (green) outline
are the values corrected for dust depletion. No dust correction is 
required for Q0454+039 where [Zn/Cr]~$\simeq 0$ within the errors. We 
have not corrected the measurements for Q1351+318 because here 
depletions are greater than a factor of two and the assumption
that all refractory elements are depleted by the same amount may break 
down.}
\end{figure}

%
%

\begin{figure}
\figurenum{7}
\vspace*{-5.5cm}
\hspace*{-2.5cm}
\psfig{figure=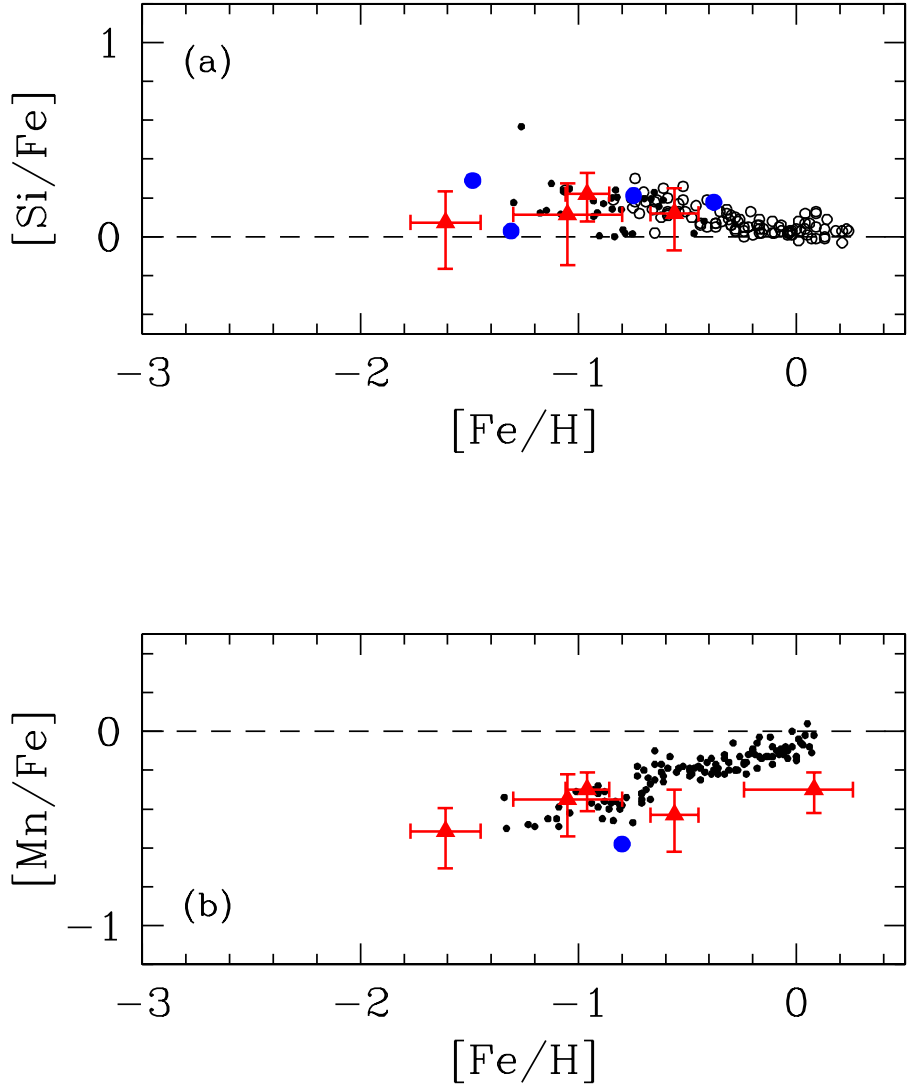,width=195mm}
\vspace{-5.5cm}
\figcaption{Metallicity dependence of the abundances of Si and Mn. 
Small dots are values in Galactic stars.
For Si we have reproduced the stellar data
from the compilations by 
Edvardsson et al. (1993; open dots) 
and Nissen \& Schuster (1997; filled dots);
to avoid overcrowding in the figure we have only plotted one out of 
every two measurements by Edvardsson et al.
For Mn we show the results of  
the recent survey by Nissen et al. (2000).
The other symbols refer to damped \lya\ systems observed by us (triangles), 
or by others (large dots); Zn has been taken as a proxy for Fe.
The DLAs shown are cases where the 
corrections for the fractions of Si and Mn locked up in grains are less 
than a factor of two.}
\end{figure}

\end{document}